\documentclass[fleqn,usenatbib]{mnras}
\usepackage{newtxtext,newtxmath}

\usepackage[T1]{fontenc}

\DeclareRobustCommand{\VAN}[3]{#2}
\let\VANthebibliography\thebibliography
\def\thebibliography{\DeclareRobustCommand{\VAN}[3]{##3}\VANthebibliography}

\usepackage{graphicx}	
\usepackage{amsmath}	
\usepackage{comment}

\title[Modelling Pop III stars]{Modelling Population III stars for semi-numerical simulations}

\author[Tanaka and Hasegawa]{
Toshiyuki Tanaka$^{1}$\thanks{E-mail: tanaka.toshiyuki@d.mbox.nagoya-u.ac.jp} and
Kenji Hasegawa$^{1}$
\\
$^{1}$Division of Particle and Astrophysical Science, Graduate School of Science, Nagoya University, Chikusa, Nagoya 464-8602, Japan\\
}

\date{Accepted XXX. Received YYY; in original form ZZZ}

\pubyear{2020}

\begin{document}
\label{firstpage}
\pagerange{\pageref{firstpage}--\pageref{lastpage}}
\maketitle

\begin{abstract}
Theoretically modelling the 21-cm signals caused by Population III stars (Pop III stars) is the key to extracting fruitful information on Pop III stars from current and forthcoming 21-cm observations. In this work we develop a new module of Pop III stars in which the escape fractions of ionizing photons and Lyman-Werner (LW) photons, photo-heating by UV radiation, and LW feedback are consistently incorporated. 
By implementing the module into a public 21-cm semi-numerical simulation code, {\small 21CMFAST}, we demonstrate 21-cm signal calculations and investigate the importance of Pop III star modelling.
What we find is that the contribution from Pop III stars to cosmic reionization significantly depends on the treatment of the escape fraction. 
With our escape fraction model, Pop III stars hardly contribute to reionization because less massive halos, whose escape fraction are high, cannot host Pop III stars due to LW feedback. On the other hand, Pop III stars well contribute to reionization with the conventional constant escape fraction.   
We also find that UV photo-heating has non-negligible impact on the 21-cm global signal and the 21-cm power spectrum if the ionization fraction of the Universe is higher than roughly 1 percent. 
In this case, the strength of the 21-cm global signal depends on the photo-heating efficiency and thus on the Pop III star mass. 
We conclude that detailed modelling of Pop III stars is imperative to predict 21-cm observables accurately for future observations.
\end{abstract}




\section{Introduction}
At the end of the cosmic dark ages, the first generation of stars have formed in mini-haloes (MHs) whose mass is $\sim 10^{5-8} \mathrm{M}_\odot$\citep[e.g.][]{Tegmark1997, Nishi1999, Abel2002, Bromm2002, Gao2007, O'Shea2007, Yoshida2008}. 
Such zero-metal stars formed from the primordial gas is called Population III stars (Pop III stars). Because of the low virial temperature of MHs ($T_\mathrm{vir} \lesssim 10^4$ K), the main coolant is molecular hydrogen whose relatively ineffective cooling leads to high gas accretion rates on protostars. 
Therefore, Pop III stars are expected to be massive ($\sim 10-1000 \mathrm{M}_\odot$). 
By reason of the importance of molecular hydrogen as a coolant, the dissociation photons in the Lyman-Werner band (LW photons; 10.2-13.6 eV) work as negative feedback on Pop III star formation \citep[e.g.][]{Haiman1997, Susa2006, Hasegawa2009}.

Pop III stars have key roles in the history of the Universe:
Pop III stars are one of the candidates which provide the intergalactic medium (IGM) with ionizing photons, contributing to the beginning of cosmic reionization \citep{Alvarez2006, Johnson2007}.
Also, since Pop III stars are thought to be massive, they would cause supernova at the end of its lifetime, leaving black holes \citep{Venemans2013, Banados2014, Wu2015} and providing metals which enhance the subsequent star formation \citep{Wise2012, Karlsson2013}.
How large the Pop III stars impact on the reionization history and the structure formation in the Universe strongly depends on their properties.
Therefore, several theoretical works have studied the properties such as stellar mass spectrum \citep{Susa2014, Hirano2015} and star formation rate density \citep{Wise2012, Visbal2020}. 
However, results of these works have not converged yet.

From the observational point of view, 21-cm line observations are expected to provide valuable data to understand the properties of Pop III stars, because the signal reflects the state of gas such as the ionization fraction and the spin temperature of the hyper-fine structure of neutral hydrogen.
Current observatories (e.g. the Murchison Widefield Array, the Low Frequency Array) have already put upper limits on the power spectrum of the 21-cm brightness temperature \citep{Paciga2013, Dillon2014, Dillon2015, Ewall-Wice2016, Beardsley2016, Patil2017, Gehlot2019, Eastwood2019, Kolopanis2019, Barry2019}. 
Also the Experiment to Detect the Global Epoch of Reionization Signature (EDGES) has reported the absorption signal at frequency 78 MHz which corresponds to $z\sim17$ \citep{Bowman2018}.
The forthcoming instrument, the Square Kilometre Array (SKA), is about to start its observation.

In order to extract information on Pop III stars from observed 21-cm signals, we need to theoretically model the 21-cm signal originated in Pop III stars accurately. 
\cite{Ahn2012} have conducted the radiative transfer (RT) simulations with the sub-grid model about the number of MHs in simulation grids, instead of directly resolving MHs which need enormous computational resources. 
They have calculated the 21-cm power spectrum assuming the tight coupling between the gas kinetic temperature and the spin temperature \citep{Shapiro2012}. 
\cite{Visbal2020} have used semi-analytic methods to efficiently calculate the formation of Pop III stars and inhomogeneous ionization field.
However, the models of Pop III stars in these previous studies are rather simple. 
For instance, they assume constant escape fraction of ionizing photons, though the escape fraction is known to depend on the halo mass and the stellar mass in the halo \citep{Kitayama2004, TT2018}. 
This fact indicates that the dependence of 21-cm signal on the properties of Pop III stars could be  stronger than previously thought. 
The LW negative feedback boosts the minimum halo mass above which Pop III star formation is possible. 
Therefore, the escape fraction of LW photons is also important to determine the cosmic star formation rate density. 

Moreover, the photo-heating by UV radiation has often been neglected in 21-cm signal simulations because its mean free path is relatively short so that the UV-heated region almost corresponds to the ionized region. 
However, \cite{Yajima2014} clearly showed that Pop III stars create large partially ionized regions due to their high effective temperature and such heated regions show 21-cm emission. 
Using the results from radiation hydrodynamics (RHD) simulations, \cite{TT2018} (hereafter TT18) also found that this UV heating has non-negligible effect on the global 21-cm signal if cosmic star formation rate density is high. 

The main purpose of this paper is to develop a new module of Pop III stars, in which the escape fractions of the ionizing and LW photons, UV-heating, and LW feedback are consistently considered. 
By implementing our module into a public semi-numerical simulation code,  {\small 21CMFAST} \citep{Mesinger2011}, we also demonstrate 21-cm signal calculations and discuss the importance of Pop III star modelling. 

This paper is organized as follows. 
In Section~\ref{sec:method}, we describe the method to incorporate the escape fractions, UV heating, and LW negative feedback. 
Then, showing simulation results in Section~\ref{sec:results}, we explain how our model affects the ionization state and temperature of the IGM, the 21-cm global signal, and the 21-cm power spectrum.
We next discuss the limitation and uncertainties of this work in Section~\ref{sec:discussion}, and finally make conclusions in Section~\ref{sec:conclusion}.

Throughout this paper, we work with a flat $\Lambda$CDM cosmology: 
the matter density, $\Omega_\mathrm{m0} = 0.308$, the baryon density, $\Omega_\mathrm{b0}=0.0485$, the Hubble constant, $h_\mathrm{0} = 0.678$, the index of initial matter power spectrum, $n_\mathrm{s} = 0.968$, the rms mass fluctuation on 8 Mpc$/h$ scales at present day, $\sigma_8 = 0.828$, the mass fraction of Helium, $Y_\mathrm{He} = 0.249$ (Planck Collaboration XIII \citeyear{PlanckXIII}).

\section{Method}
\label{sec:method}
In this section, we first summarize the basics of the 21-cm signal, then, describe the modification of the ionization calculation of {\small 21CMFAST} and the methods to take into account the escape fractions, UV heating, and  LW feedback. 

The cosmological 21-cm signal can be written as \citep[e.g.][]{Furlanetto2006Oh}
\begin{equation}
    \label{eq:Tb}
    \delta T_\mathrm{b} = 
    38.7 x_\mathrm{HI} (1+\delta) \left(\frac{1+z}{20}\right)^{1/2}
    \frac{T_\mathrm{S}-T_\mathrm{CMB}(z)}{T_\mathrm{S}}
    \mathrm{mK},
\end{equation}
where $x_\mathrm{HI}$ is fraction of neutral hydrogen, $\delta$ is overdensity, and $T_\mathrm{CMB}(z)$ is the CMB temperature at redshift $z$, respectively. The spin temperature $T_\mathrm{S}$ is determined by
\begin{equation}
    \label{eq:Ts}
    T_\mathrm{S}^{-1}=
    \frac{T_\mathrm{CMB}^{-1} + x_\mathrm{c} T_\mathrm{gas}^{-1} + x_\mathrm{\alpha} T_\mathrm{\alpha}^{-1}
    }{1+ x_\mathrm{c}+ x_\mathrm{\alpha}},
\end{equation}
where $T_\mathrm{gas}$, $T_\mathrm{\alpha}$, $x_\mathrm{c}$, and $x_\mathrm{\alpha}$ are the gas kinetic temperature, the color temperature of Ly$\alpha$ photons, the collisional coupling coefficient, and the Ly$\alpha$ coupling coefficient, respectively. 
We assume $T_\mathrm{\alpha} = T_\mathrm{gas}$ because of the large cross section of Ly$\alpha$ photons to neural hydrogen.

We use  {\small 21CMFAST} \citep{Mesinger2011} with our new Pop III star model. 
The initial condition is generated in roughly the same way as ordinary cosmological simulations.
The density field is extrapolated from the initial condition using the first order perturbation theory with the correction of the second-order Lagrangian perturbation theory. 
The  {\small 21CMFAST} treats inhomogeneous Ly$\alpha$ radiation field as well as ionizing radiation field. 
The emissivity of ionizing photons is controlled by the parameter, $N_\mathrm{UV}$ which is the number of ionizing photons produced by single stellar baryon, on the other hand, the emisivity of Lyman series photons is based on the spectral model of \cite{Barkana2005}.
The self-consistently generated 3D distributions of density, HI fraction, and spin temperature are used to obtain the 21-cm brightness temperature field.

\subsection{Ionization field}
\label{sec:sub_ion_field}

\subsubsection{Ionization calculation including recombination} 
We here describe how we calculate the ionization field. 
In the original  {\small 21CMFAST}, the criterion for judging whether the grid is ionized or not is,
\begin{equation}
    \label{eq:ionjudge_original}
   \zeta_\mathrm{ion} \bar{f}_\mathrm{coll}^{R} > 1 + \bar{n}_\mathrm{rec}^{R},
\end{equation}
where $\zeta_\mathrm{ion}=N_\mathrm{UV} f_\mathrm{esc} f_*$ with being
$f_\mathrm{esc}$ the escape fraction of ionizing photons and $f_*$ star formation efficiency,
$\bar{f}_\mathrm{coll}^{R}$ is the collapse fraction spatially averaged over scale $R$,
and $\bar{n}_\mathrm{rec}^R$ is the average number of recombination per baryon on scale $R$.

We adopt a new ionization criterion which is slightly different from that in the original code,
\begin{equation}
    \label{eq:ion_criterion}
    \bar{N}_\mathrm{ion}^R > 1 + \bar{N}_\mathrm{rec}^R,
\end{equation}
where $\bar{N}_\mathrm{ion} ^R$ and  $\bar{N}_\mathrm{rec} ^R$ are the numbers of ionizing photons and recombination per baryon spatially averaged over a certain scale $R$, based on values at each grid, ${N}_\mathrm{ion}(\mathbf{x})$ and ${N}_\mathrm{rec}(\mathbf{x})$:
\begin{equation}
    N_\mathrm{ion} =
    \int_z^{z_\mathrm{init}} \mathrm{d} z\
    \zeta_\mathrm{ion}(z)
    \frac{\mathrm{d}f_\mathrm{coll}(z)}{\mathrm{d}z},
\end{equation}
\begin{equation}
    \label{eq:Nrec}
    N_\mathrm{rec} =  \int_z^{z_\mathrm{init}} \mathrm{d} z\ \alpha_\mathrm{B} n_\mathrm{H}
    (N_\mathrm{ion,prev} - N_\mathrm{rec,prev})
    \frac{\mathrm{d}t}{\mathrm{d}z},
\end{equation}
where, $\alpha_\mathrm{B}$ is the case-B recombination coefficient,
$n_\mathrm{H}$ is number density of hydrogen, $N_\mathrm{ion,prev}$ and $N_\mathrm{rec,prev}$ are the values at the previous time step. 
Note that $\bar{N}_\mathrm{rec}^R$ is different from $\bar{n}_\mathrm{rec}^R$ introduced by \cite{Sobacchi2014}: The $\bar{n}_\mathrm{rec}^R$ is the recombination number in a fully ionized cell to keep the grid ionized, which means they do not take into account recombination before the cell is fully ionized. 
On the other hand, our $\bar{N}_\mathrm{rec}^R$ additionally includes the recombination when the cell is partially ionized. 
Indeed \cite{Visbal2018} have showed that the recombination we additionally include can be ignored due to the rapid growth of star formation rate density. 
However, with our scheme, the escape fraction would drop to nearly zero so that the emissivity of ionizing photons can decrease. 
Therefore, the recombination is needed to be taken into account.

With the criterion,  Equation~(\ref{eq:ion_criterion}), simulation cells are judged whether they are fully ionized or not from the largest scale $R_\mathrm{max}$ to the cell scale.
We take $30~\mathrm{Mpc}$ as $R_\mathrm{max}$ which is large enough in the high-$z$ Universe we are interested in.
If the cells are not judged as fully ionized even at the cell scale, the ionization fraction is computed as $x_\mathrm{e} =$ min\{$N_\mathrm{ion} - N_\mathrm{rec}$, 1\}. 

As in  {\small 21CMFAST}, we assume that hydrogen and single ionized helium have the same ionization fraction $x_e$. 
Ignoring helium recombination in Equation (\ref{eq:Nrec}) is indeed good approximation given that photons produced by helium recombination have enough energy to ionize hydrogen.

We have confirmed that our algorithm can solve Str\"{o}mgren sphere with errors less than $\sim 10\%$.

\subsubsection{Escape fraction of ionizing photons}
\label{subsubsec:fesc}
\begin{figure}
    \includegraphics[width=\columnwidth]{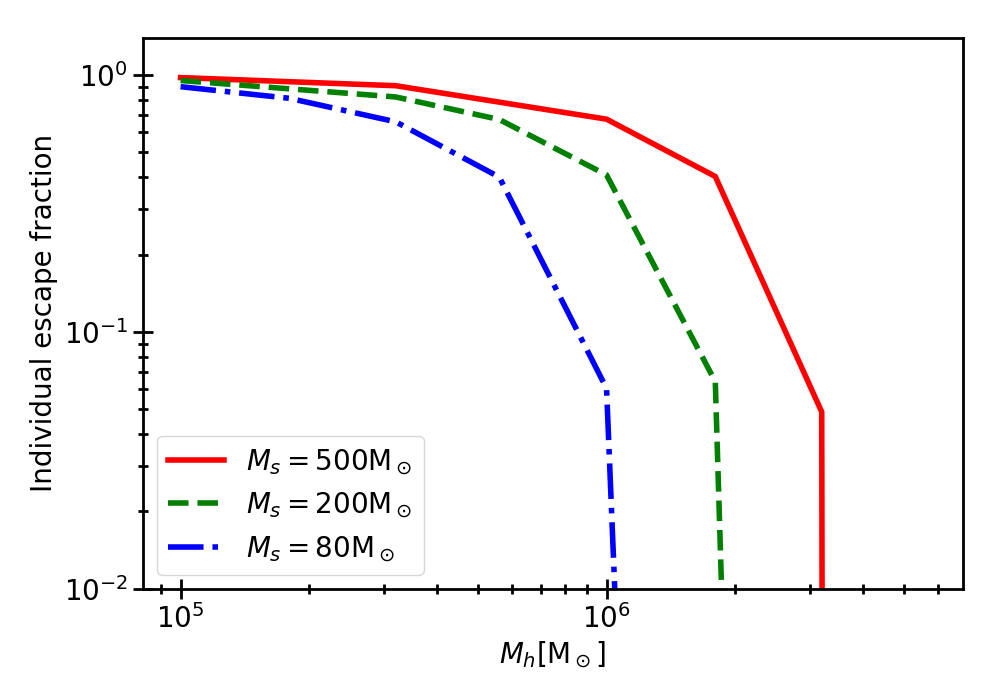}
    \caption{The escape fraction of individual halo $\mathscr{f}_\mathrm{esc}$ as a function of halo mass at redshift 20. The red solid line, the green dashed line, and blue dashed-dotted line are the cases of the stellar mass of $M_\mathrm{star} = 500$M$_\odot$, $200$M$_\odot$, and $80$M$_\odot$, respectively.
    }
    \label{fig:infesc}
\end{figure}
In our model, we consider the escape fraction depending on halo mass and stellar mass.
Using the escape fraction of ionizing photons for an individual halo hosting a Pop III star $\mathscr{f}_\mathrm{esc}(M_\mathrm{halo},M_\mathrm{star})$, the averaged escape fraction at a given redshift $z$ can be written as 
\begin{equation}
    \label{eq:fesc}
    f_\mathrm{esc} (z) =
    \frac{\int_{M_\mathrm{cool}}^{\infty} \mathrm{d}M_\mathrm{halo} \frac{\mathrm{d}n}{\mathrm{d}M_\mathrm{halo}} \mathscr{f}_\mathrm{esc}(M_\mathrm{halo},M_\mathrm{star})}{\int_{M_\mathrm{cool}}^\infty \mathrm{d}M_\mathrm{halo} \frac{\mathrm{d}n}{\mathrm{d}M_\mathrm{halo}}}.
\end{equation}
The halo mass function (HMF), $\frac{\mathrm{d}n}{\mathrm{d}M_\mathrm{halo}}$, is obtained from a high-resolution N-body simulations \citep{Ishiyama2016} in which $4096^3$ dark matter particles are used in a $(16\mathrm{Mpc}/h)^3$ box. The minimum FoF halo mass corresponds to $1.6 \times 10^5 \mathrm{M}_\odot/h$.
Indeed the individual escape fraction depends on redshift, but the dependence is so small compared to dependences of stellar mass and halo mass that we ignore the redshift dependence in this work.

In order to investigate $\mathscr{f}_\mathrm{esc} (M_\mathrm{halo},M_\mathrm{star})$, we conduct one-dimensional spherically-symmetric RHD simulations as in TT18. 
The RHD simulation can produce the profile of 21-cm brightness temperature around an individual MH hosting a Pop III star by solving gas dynamics, radiative transfer of ionizing photons, and non-equilibrium chemical reactions. 
In this RHD simulations, we assume that one MH hosts one PopIII star at the centre. 

Using the RHD simulation results, the escape fraction at $t_\mathrm{age}$ is calculated as
\begin{equation}
    \tilde{\mathscr{f}}_\mathrm{esc} (t_\mathrm{age}) =
    \frac{\int_{\nu_{L}}^\infty \mathrm{d}\nu \frac{L_\nu}{h\nu}\exp\{{-\tau_\nu(r_\mathrm{vir})}\}}{\int_{\nu_{L}}^\infty \mathrm{d}\nu \frac{L_\nu}{h\nu}}
\end{equation}
where $L_\nu$ is specific luminosity of Pop III stars, $\nu_\mathrm{L}$ is Lyman limit frequency, $r_\mathrm{vir}$ is the virial radius of the halo. 
We approximate the stellar spectrum as the black-body spectrum with the effective temperature of Pop III stars \citep{Schaerer2002}.
Then, averaging $\tilde{\mathscr{f}}_\mathrm{esc}$ over stellar lifetime, $t_\mathrm{life}$, gives $\mathscr{f}_\mathrm{esc} = \int_0^{t_\mathrm{life}} \tilde{\mathscr{f}}_\mathrm{esc} \mathrm{d}t_\mathrm{age} / t_\mathrm{life}$.

\begin{figure}
    \includegraphics[width=\columnwidth]{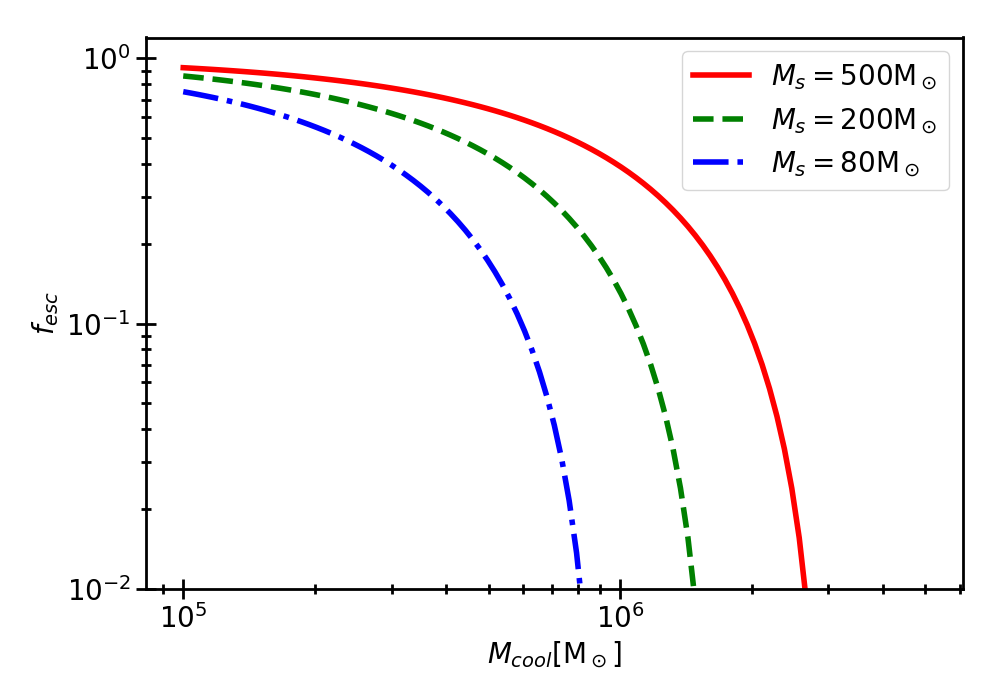}
    \caption{The averaged escape fraction as a function of the minimum halo mass for star formation, $M_\mathrm{cool}$. The red solid, the green dashed, and the blue dashed-dotted lines are the cases of the stellar mass of $M_\mathrm{star} =500$M$_\odot$, $200$M$_\odot$, and $80$M$_\odot$, respectively.
    }
    \label{fig:fescMcool}
\end{figure}

In TT18, the behavior of $\tilde{\mathscr{f}}_\mathrm{esc}$ is described in detail. 
Thus, we here shows the halo mass dependence of the individual escape fraction in Fig.~\ref{fig:infesc}.
In the case of less massive haloes, the ionized bubble easily expand beyond the virial radius due to less amount of gas inside. 
Therefore, soon after the Pop III star is born, the $\tilde{\mathscr{f}}_\mathrm{esc}$ become almost unity, resulting in large $\mathscr{f}_\mathrm{esc}$.
As the halo mass increases, the gas tends to prevent the ionization front from expanding so that the resultant $\tilde{\mathscr{f}}_\mathrm{esc}$ is small. 
Thus, in the cases of massive MHs ($M_\mathrm{h}\gtrsim$ a few $\times 10^6$M$_\odot$), the $\tilde{\mathscr{f}}_\mathrm{esc}$ remains nearly zero even at the end of the lifetime. 
Consequently the individual escape fraction ${\mathscr{f}}_\mathrm{esc}$ is almost zero.
By simulating in wide range of parameter space, we develop useful fitting formula for $\mathscr{f}_\mathrm{esc}$
\begin{equation}
    \mathscr{f}_\mathrm{esc} = \mathrm{MAX}\left\{
    -18.14 M_\mathrm{star}^{-0.67} \left( \frac{M_\mathrm{halo}}{10^6[\mathrm{M}_\odot]} \right) + 0.97,\ 0 \, \right\}.
\end{equation}
The formula is valid with $M_\mathrm{halo} > 10^5 \mathrm{M}_\odot$ and $M_\mathrm{star} = [40 - 500] \mathrm{M}_\odot $. 
As described in later section, we apply this escape fraction model not only to ionizing radiation but to LW radiation. 

Finally, equation \eqref{eq:fesc} can be solved once $M_\mathrm{cool}$ is determined. 
We show the escape fraction $f_\mathrm{esc}$ as a function of $M_\mathrm{cool}$ in Fig.~\ref{fig:fescMcool}. 
As $M_\mathrm{cool}$ rises, the escape fraction falls. This is because the individual escape fraction ${\mathscr{f}}_\mathrm{esc}$ is smaller for more massive haloes. 
Since $M_\mathrm{cool}$ is determined by the LW background intensity controlled by the escape fraction of LW photons, we solve the redshift evolution of the escape fractions and $M_\mathrm{cool}$, consistently.

\subsection{Photo-heating by UV radiation}
\label{subsec:3fluid}
Even though the mean free path of UV radiation emitted from massive PopIII stars is longer than that of galaxies and have considerable impact on surrounding 21-cm signature, it is still too short \footnote{In the case of X-ray, mean free path is exceedingly long ($\sim$ 1 Gpc at $z=20$, $h\nu = 5$ keV, assuming cosmic mean baryon density) so that X-ray heating can be treated as uniform heating in grid. 
On the other hand, the mean free path of UV photons ($\sim$ 1 kpc at $z=20$, $h\nu = 50$ eV) is much shorter than the size of a simulation grid.} compared to a simulation grid size which is typically $\sim 1$ comoving Mpc. 

In order to incorporate the inhomogeneous UV heating in sub-grid scale, we divide the whole region of a grid into three different regions which are ionized, cold, and heated regions. 
The ionized region is where no neutral hydrogen exist and shows no 21-cm signal. 
The cold region is neutral region and not heated so that the gas in this region absorb the background CMB radiation. 
The heated region is slightly ionized and its gas temperature is higher than the CMB temperature, creating emission signal. 
When we write the 21-cm brightness temperatures in these three regions as $\delta T_\mathrm{b,ion}$, $\delta T_\mathrm{b,cold}$, and $\delta T_\mathrm{b,heat}$, the 21-cm signal of the grid can be written as
\begin{equation}
    \delta T_\mathrm{b,grid} = \sum_i f_i
    \delta T_{\mathrm{b},i} \ \ \
    (i=\mathrm{ion},\, \mathrm{cold},\, \mathrm{heat}),
\end{equation}
where $f_i$ is volume fraction of $i$ region. 
Since $\sum_i f_i = 1$, we need ionized fraction $f_\mathrm{ion}$ and ratio of heated region to ionized region, $\gamma_\mathrm{h/i} \equiv f_\mathrm{heat}/f_\mathrm{ion}$, where $f_\mathrm{ion}$ can be obtained by the ionization field calculation described in \ref{sec:sub_ion_field}.
The 21-cm brightness temperatures in these three regions are expressed as
\begin{equation}
    \delta T_\mathrm{b,ion} = 0 \ \mathrm{mK},
\end{equation}
\begin{equation}
    \delta T_\mathrm{b, cold}= 38.7 (1+\delta)
    \left(\frac{1+z}{20}\right)^{1/2}
    \frac{T_\mathrm{S}-T_\mathrm{CMB}(z)}{T_\mathrm{S}} \ \mathrm{mK},
\end{equation}    
\begin{equation}
    \label{eq:Tbheat}
    \delta T_\mathrm{b, heat}= 38.7 (1+\delta)
    \left(\frac{1+z}{20}\right)^{1/2} \ \mathrm{mK},
\end{equation}
where we assume strong Ly$\alpha$ coupling in the heated region because the region is close enough to Pop III stars to have high intensity of Lyman series photons as shown by TT18.

The RHD simulations described in Section \ref{subsubsec:fesc} also gives $\gamma_\mathrm{h/i}$. 
As some studies have shown \citep[TT18]{Chen2008, Yajima2014}, the 21-cm signature around Pop III star has a typical structure: 
The most-inner region is ionized region where no neutral hydrogen remain, resulting in zero signal. 
The next outer region is heated region where gas is partially ionized and hotter than the CMB temperature so that the emerging signal is emission. 
The region outside the heated region is the cold region where gas remain cold below the CMB temperature while the Ly$\alpha$ coupling is sufficiently strong to couple the spin temperature with the cold gas temperature, showing the absorption signal.

To estimate $\gamma_\mathrm{h/i}$, we use the profiles of ionized fraction and 21-cm brightness temperature calculated with the RHD simulations. 
We locate ionized region and heated region around MHs as follows: The boundary between ionized region and heated region is defined as the shell whose neutral fraction is 1\%. Whereas the boundary between heated region and cold region is defined as where $\delta T_\mathrm{b}$ first turns to be positive, judging from the most-outer cold shell. 
Once we locate the radius of ionized region $r_\mathrm{ion}$ and the outer radius of heated region $r_\mathrm{heat}$, $\gamma_\mathrm{h/i}$ can be calculated as $\gamma_\mathrm{h/i} = (r_\mathrm{heat}^3 - r_\mathrm{ion}^3)/ r_\mathrm{ion}^3$. 
The ionization fraction at heated region defined with this way is less than 10\%, therefore we treat heated region as neutral region for simplicity.

\begin{figure}
	\includegraphics[width=\columnwidth]{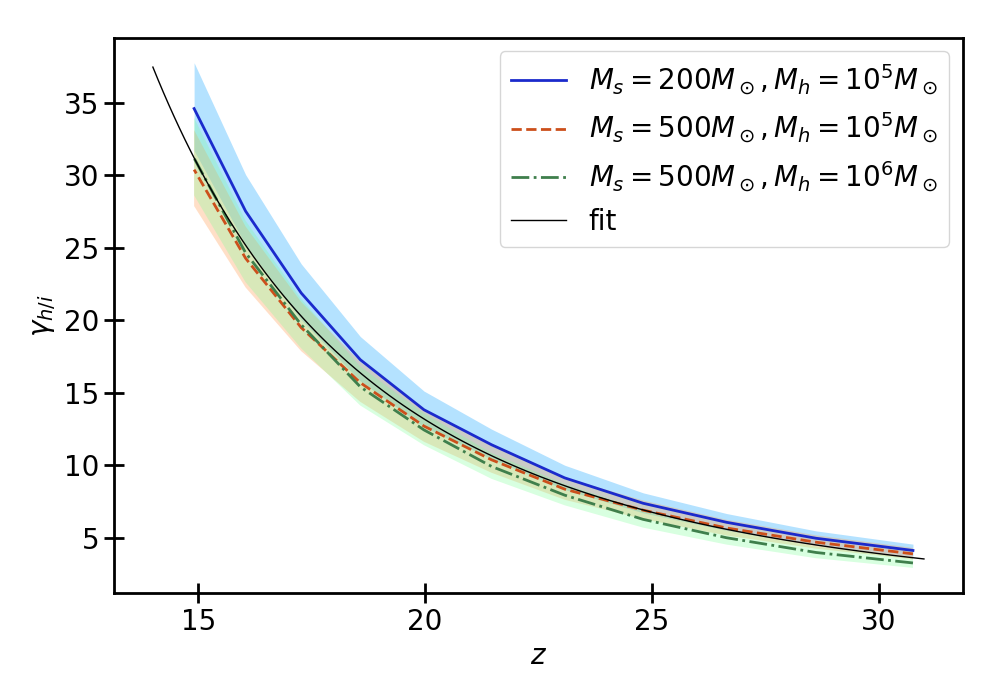}
    \caption{
    $\gamma_\mathrm{h/i}$ as a function of redshift.
    The blue solid line is the case of $M_\mathrm{star} = 200\mathrm{M}_\odot, M_\mathrm{halo} = 10^5\mathrm{M}_\odot$, the orange dashed line is the case of $M_\mathrm{star} = 500\mathrm{M}_\odot, M_\mathrm{halo} = 10^5\mathrm{M}_\odot$,  the green dashed-dotted line is the case of $M_\mathrm{star} = 500\mathrm{M}_\odot, M_\mathrm{halo} = 10^6\mathrm{M}_\odot$. The shaded regions represent the errors originating from the spacial resolution of the one-dimension RHD simulations. The thin black solid line is fitted line. }
    \label{fig:gamz}
\end{figure}

Fig.~\ref{fig:gamz} shows $\gamma_\mathrm{h/i}$ as a function of redshift for some cases. 
In all the cases, $\gamma_\mathrm{h/i}$ increases with time. 
To understand the behaviour, we write the ratio as $\gamma_\mathrm{h/i} = 3(l/r_\mathrm{ion}) + 3(l/r_\mathrm{ion})^2 + 3(l/r_\mathrm{ion})^3 $ where $l \equiv r_\mathrm{heat} - r_\mathrm{ion}$ is the width of heated region which roughly corresponds to the average mean free path of UV photons. 
To think simply, we regard ionized region as ionized sphere which located at uniform density field and its ionization status is far from the ionization equilibrium. Then the radius of ionized region obeys $r_\mathrm{ion} \propto n^{-1/3}$ because of $N_\mathrm{ion} = (4/3)\pi r_\mathrm{ion}^3 n$ for given ionizing photon number $N_\mathrm{ion}$, where $n$ is number density of gas, assuming the gas is composed of only hydrogen. 
On the other hand, $l \propto n^{-1}$, therefore as redshift increases and the mean density of the Universe gets denser, $l/r_\mathrm{ion}$ becomes smaller, resulting in smaller $\gamma_\mathrm{h/i}$.

The effective temperature indeed depends on the stellar mass so that $\gamma_\mathrm{h/i}$ is expected to depend on the stellar mass. 
However the stellar mass dependence of $\gamma_\mathrm{h/i}$ is found to be smaller than the error which is originated from the spacial resolution of the RHD simulations (See the blue solid line, orange dashed line, and these error regions in Fig.~\ref{fig:gamz}). 
Hence we neglect the stellar mass dependence of $\gamma_\mathrm{h/i}$. 

The halo mass dependence is too small to affect the  $\gamma_\mathrm{h/i}$ (See the dashed orange line and the green dash-dotted line in Fig.~\ref{fig:gamz}) as well as the stellar mass dependence. 
Thus, we neglect the dependence in this work. 
Although $\gamma_\mathrm{h/i}$ of massive MHs ($\gtrsim 3\times10^6 \mathrm{M}_\odot$) would be different because of the dense gas profile in and near the vrial radius, Pop III stars in such massive haloes do not emit enough ionizing photons into the IGM due to significantly small escape as shown in Section~\ref{subsubsec:fesc}. 

We develop the fitting formula for  $\gamma_\mathrm{h/i}$ (the thin black solid line in Fig.~\ref{fig:gamz}):
\begin{equation}
    \log(\gamma_\mathrm{h/i}) =
    -3.11\log(1+z) + 5.23.
\end{equation}
We use the formula to obtain the value of $\gamma_\mathrm{h/i}$.

\subsection{LW feedback}
LW photons dissociate molecular hydrogen, suppressing the formation of Pop III stars in MHs.
In this work, we take into account the negative feedback and its redshift evolution by calculating the box-averaged LW intensity at each redshift.

The minimum halo mass above which Pop III stars can form, $M_\mathrm{cool}$, depends on the LW intensity, $J_\mathrm{LW}$. 
The relation of $J_\mathrm{LW}$ and $M_\mathrm{cool}$ is investigated with numerical simulations \citep{Machacek2001, Wise2007, OShea2008}, and it is well fitted by \citep{Visbal2014}
\begin{align}
    M_\mathrm{cool}=
    3.4 \times 10^5 
    \left( \frac{1+z}{21} \right)^{-1.5}
    \times \left\{1+6.96
    \left(4\pi J_\mathrm{LW}(z)\right)^{0.47}
    \right\} M_\odot,
\end{align}
where, the unit of $J_\mathrm{LW}$ is [10$^{-21}$ erg s$^{-1}$ cm$^{-2}$ Hz$^{-1}$ str$^{-1}$]. 

We calculate  $J_\mathrm{LW}(z)$ by averaging the LW intensity at each grid,  $\mathcal{J}_\mathrm{LW}(\mathbf{x},z)$, over the simulation box. $\mathcal{J}_\mathrm{LW}(\mathbf{x},z)$ is obtained by the way similar to that of X-ray intensity implemented in the original  {\small 21CMFAST} and that used in \cite{Fialkov2013}. 
We sum up contributions from shells whose centre is at the grid position $\mathbf{x}$
\begin{equation}
    \label{eq:jlw}
    \mathcal{J}_\mathrm{LW}(\mathbf{x},z) =
    \int_z^{z_\textrm{max}} \mathrm{d}z' \frac{1}{4\pi} \frac{1}{4\pi r_\mathrm{p}^2} \frac{\mathrm{d}\varepsilon(\textbf{x},z')}{\mathrm{d}z'},
\end{equation}
where $z'$ is corresponding to the redshift when the reached photons to the grid is emitted in each shell, and the $r_\mathrm{p}$ is the proper distance from the shell.  
The LW specific emissivity per redshift can be written as
\begin{equation}
    \label{eq:epslw}
    \frac{\mathrm{d}\varepsilon(\textbf{x},z')}{\mathrm{d}z'}=
    \left( \frac{N_\mathrm{LW}E_\mathrm{LW}}{\Delta\nu_\mathrm{LW}}\right)
    f_* f_\mathrm{esc,LW}\, \bar{n}_\mathrm{b,0} (1+\bar{\delta}_R)
    \frac{\mathrm{d}V}{\mathrm{d}z'}
    \frac{\mathrm{d}f_\mathrm{coll}}{\mathrm{d}t},
\end{equation}
where $f_\mathrm{esc,LW}$ is the escape fraction of LW photons, $\bar{n}_\mathrm{b,0}$ is the mean baryon number density at $z=0$, $\bar{\delta}_R$ is the overdensity averaged over the shell scale $R$, $\mathrm{d}V$ is the comoving volume of the shell, and $\mathrm{d}z$ is redshift width corresponding to the shell width, respectively. 
The factor $(N_\mathrm{LW}E_\mathrm{LW}/{\Delta\nu_\mathrm{LW}})$ means the energy of LW photons per stellar baryon per frequency. 
We adopt values used in \cite{Mebane2018}.
The escape fraction of dissociation photons, $f_\mathrm{esc,LW}$, has similar value with that of ionizing photons \citep{Kitayama2004}. 
Therefore, we assume that the escape fractions of LW and ionizing photons are identical for simplicity (see ~\ref{sec:sub_ion_field}). 
In terms of $z_\mathrm{max}$, we assume all LW band photons redshift by 4\% before absorbed by any Lyman lines \citep{Visbal2014}
\begin{equation}
    \frac{1+z_\mathrm{max}}{1+z} = 1.04.
\end{equation}
Using equation~\eqref{eq:jlw} and \eqref{eq:epslw}, we get (c.f. equation 25 in \citealt{Mesinger2011})
\begin{align}
\begin{aligned}
    \mathcal{J}_\mathrm{LW}(\textbf{x},z) =
    \frac{f_* f_\mathrm{esc,LW}\,  \bar{n}_\mathrm{b,0}c}{4\pi} & 
    \frac{N_\mathrm{LW}E_\mathrm{LW}}{\Delta\nu_\mathrm{LW}}\\
    \times \int_z^{z_\mathrm{max}}& \mathrm{d}z' (1+z')^3
    (1+\bar{\delta}_{R})
    \frac{\mathrm{d}f_\mathrm{coll}}{\mathrm{d}z'},
\end{aligned}
\end{align}
where, $c$ is the speed of light.

\section{Demonstrations}
\label{sec:results}
In this section, we compute 21-cm signals by installing our Pop III model into  {\small 21CMFAST}, and discuss the impacts of UV heating and the escape fractions on the signal. 
All the results in this paper are from simulations with box size of (512 Mpc)$^3$, the number of grids 512$^3$. 
The initial condition are calculated with 1536$^3$ grids at redshift $300$.
The simulations start from $z = 60$ when star formation rate density of Pop III stars is too low to affect the ionization fraction of the Universe in cosmological scale. 
Since we focus on the effects by UV radiation in this work, we turn off the X-ray heating. 

The parameters used in our simulations are $N_\mathrm{UV} = 70000$, $f_* = 0.015$. With $f_* = 0.015$, resulting number of stars in MHs is less than order unity at maximum.
As for Ly$\alpha$ coupling, we use Pop III spectral models \citep{Barkana2005} prepared in the original  {\small 21CMFAST} to calculate Ly$\alpha$ intensity. 

We have conducted three simulations with our model with stellar mass of $M_\mathrm{star} = 500\mathrm{M}_\odot$, $200\mathrm{M}_\odot$, and $80\mathrm{M}_\odot$, which are respectively named as Run-Ms500, Run-Ms200, and Run-Ms80.
In order to compare our model with the conventional constant escape fraction, we also perform a simulation with the constant escape fraction of $f_\mathrm{esc} = 0.5$, named as Run-Fesc05.

\subsection{Ionization history}
\label{sec:result_ion}
\begin{figure}
    \includegraphics[width=\columnwidth]{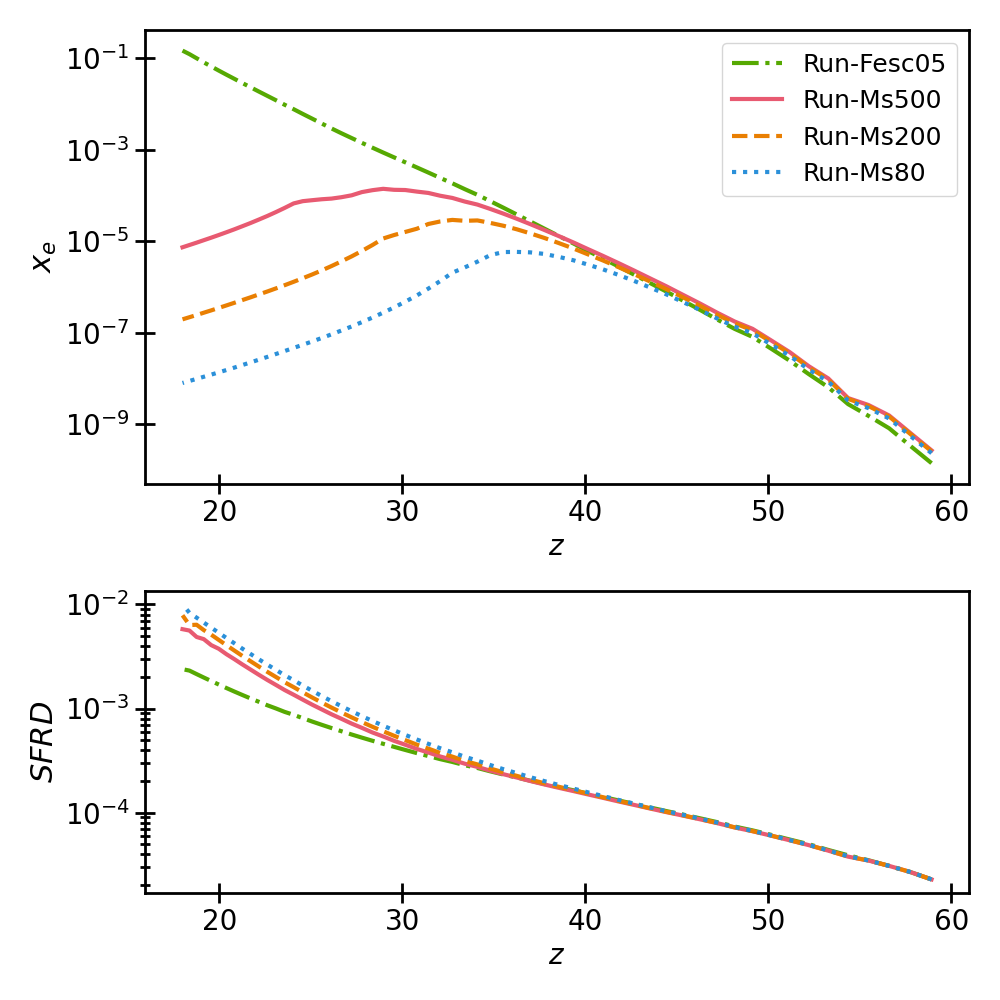}
    \caption{The box-averaged ionization fraction (top) and star formation rate density (bottom) as functions of redshift. The red solid, the orange dashed, the blue dotted, and the green dashed-dotted curves are the cases of Run-Ms500, Run-Ms200, Run-Ms80, and Run-Fesc05, respectively.
    }
    \label{fig:xe}
\end{figure}

\begin{figure}
    \includegraphics[width=\columnwidth]{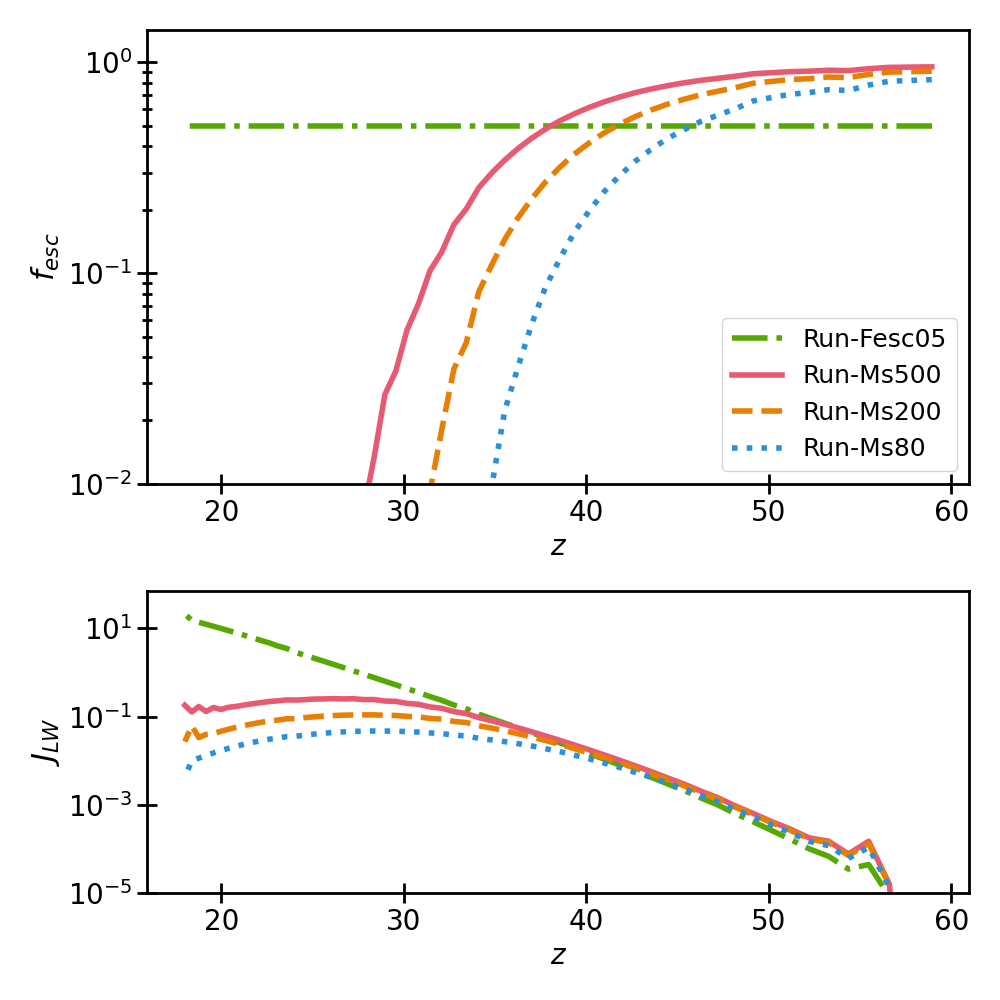}
    \caption{
    The escape fraction dependent on the LW intensity (top) and the normalized LW intensity (bottom).
    The meaning of the line types is the same as  Fig.~\ref{fig:xe}.
    }
    \label{fig:J21fesc}
\end{figure}

First of all, we show how our escape fraction model impacts on the ionization history in Fig.~\ref{fig:xe}. 
To understand the ionization history, we plot the time evolution of the escape fraction and the LW intensity in Fig.~\ref{fig:J21fesc}. 

First in the constant escape fraction model Run-Fesc05, even though the LW intensity grows with time and boosts $M_\mathrm{cool}$, more massive haloes than $M_\mathrm{cool}$ continue contributing to reionization, resulting the monotonic increase of the ionization fraction, $x_e$. 
This behaviour is qualitatively consistent with cosmological RT simulations adopting constance escape fraction \citep[e.g.][]{Ahn2012}.

With our escape fraction models (Run-Ms500, Run-Ms200, and Run-Ms80), the LW intensity is so low at high redshifts ($z\sim40-60$) that the escape fraction is nearly unity. Therefore, the box-averaged ionization fraction increases with growing star formation rate density, and is slightly larger than Run-Fesc05 owing to $f_\mathrm{esc} > 0.5$.
Then, the growth of the LW intensity gradually halts the star formation in less massive haloes. As a result, the escape fraction drops sharply at $z \sim 28 - 35$ depending on the stellar mass.
After the escape fraction drops, the LW intensity saturates. The saturated value of $J_\mathrm{LW}$ corresponds to $f_\mathrm{esc} \sim 0$. 
Consequently, the IGM is no longer irradiated with the ionizing photons from Pop III stars and the ionization fraction decreases due to the recombination. 

The peak of $x_e$ is only $\sim 10^{-4}$ even in the massive Pop III star case (Run-Ms500). The peak value is comparable with the fraction of the relic electrons. 
Thus, the Pop III stars hardly contribute to cosmic reionization.
In other words, using constant escape fraction would result in overestimation of the cosmic ionization fraction. 

\subsection{21-cm brightness temperature}
\begin{figure*}
    \includegraphics[width=1.5\columnwidth]{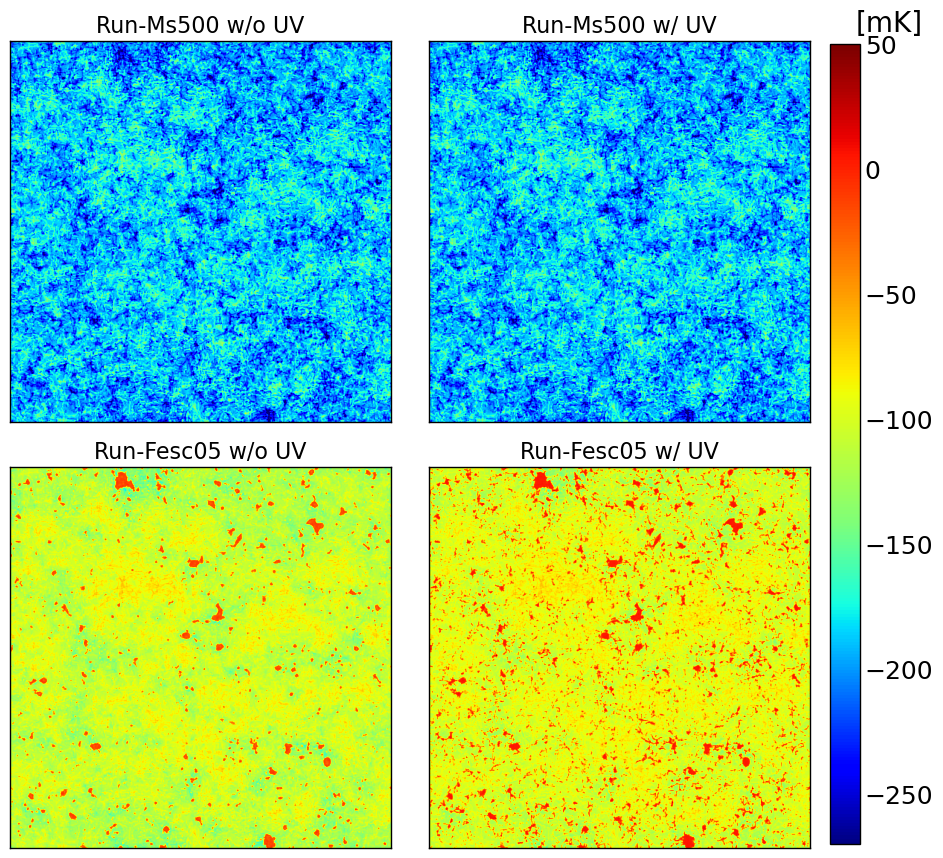}
    \caption{
    2D slices of the 21-cm brightness temperature at redshift 20. 
    The left top and the right top panels are the results of Run-Ms500 with and without UV heating. The left bottom and right bottom panels are the results of Run-Fesc05 with and without UV heating.
    Each map is 512 Mpc on a side and the thickness is 1 Mpc. The box-averaged ionization fractions are $x_e \sim 1.4 \times 10^{-5}$ and $\sim 5.3 \times 10^{-2}$ in Run-Ms500 and Run-Fesc05.
    }
    \label{fig:Tbmap}
\end{figure*}

In this section, we show the results of 21-cm brightness temperature and describe how our models influence the 21-cm signal. 

In Fig.~\ref{fig:Tbmap}, we show the 2D slices of the 21-cm brightness temperature fields at $z=20$.
Until this redshift, the intensity of Lyman series photons become moderately strong: the box-averaged coupling coefficient is $\sim$ 5.7 and 1.1 for Run-Ms500 and Run-Fesc05, respectively. 
Therefore, the spin temperature gets decoupled from the CMB temperature especially at high-density regions, showing absorption or emission signal depending on the gas kinetic temperature.
In Run-Ms500 without UV heating, since the ionization fraction is small, we see deep absorption regions at  high-density regions where $\delta$ in equation~\eqref{eq:Tb} and $x_\alpha$ in equation~\eqref{eq:Ts} are large.
The $\delta T_\mathrm{b}$ map of Run-Ms500 with UV heating looks same as that without UV heating, because the small ionization fraction means the small volume fraction of heated region.

In Run-Fesc05, on the other hand, the ionization fraction is $x_e \sim 5 \%$ so that we can see the ionized regions (red-colored regions) in the left bottom panel of Fig.~\ref{fig:Tbmap}. 
Such ionized regions corresponds to again the high-density regions and have brightness temperature closer to zero $\sim 0$~[mK], because $x_\mathrm{HI}$ in equation~\eqref{eq:Tb} is small. 
When taking into account UV heating in Run-Fesc05, $\delta T_\mathrm{b}$ is increased by the sub-grid-scale UV-heated regions. 
The increment depends on the ionization fraction of grids. As a result, the emission regions appear while the absorption signals are weakened. 
In Run-Fesc05, the LW background radiation is stronger than that in Run-Ms500. 
As a result, the star formation is significantly suppressed in Run-Fesc05 so that Ly$\alpha$ coupling is relatively weak. That is why the averaged spin temperature is higher in Run-Fesc05. 

\begin{figure}
    \includegraphics[width=\columnwidth]{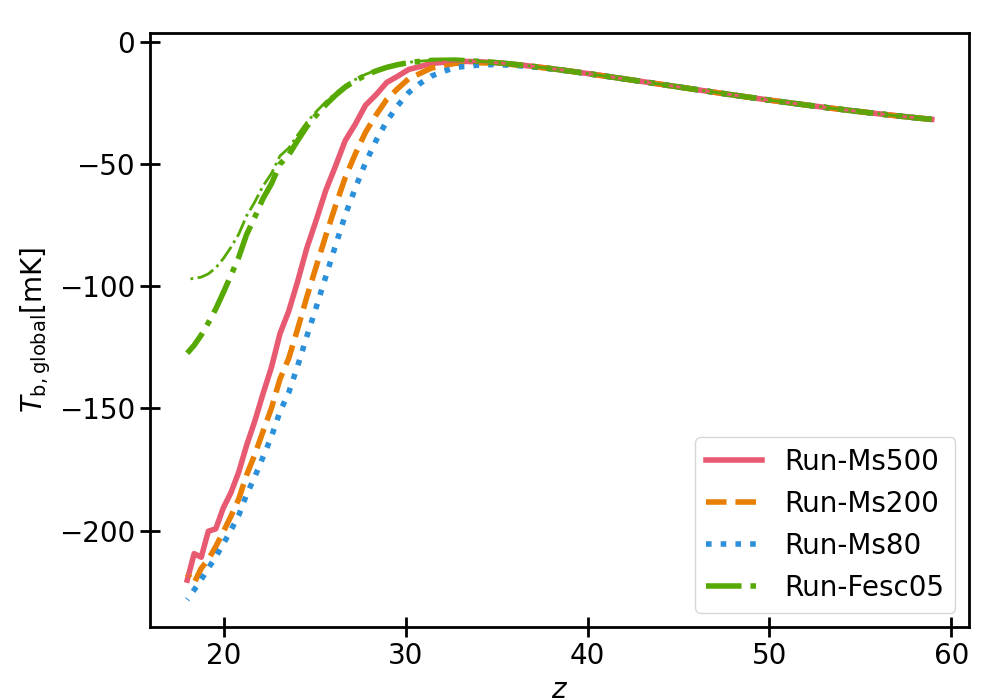}
    \caption{
    The box-averaged 21-cm brightness temperature as a function of redshift.
    The red solid, orange dashed, blue dotted, and green dashed-dotted curves are the results of Run-Ms500, Run-Ms200, Run-Ms80, and Run-Fesc05 without UV heating, respectively. 
    We also plot the results of Run-Fesc05 and Run-Ms500 with UV heating by thin lines. 
    The thin line of Run-Ms500 almost completely overlaps with Run-Ms500 without UV heating.} 
    \label{fig:Tb_global}
\end{figure}

Fig.~\ref{fig:Tb_global} shows the redshift evolution of the box-averaged 21-cm brightness temperature. 
In all the models, at $z\gtrsim 40$, the density of the Universe is so high that the spin temperature is coupled with the gas temperature via particle collisions. 
Around $z=30$, the Lyman series photons emitted from Pop III stars start to work so as to couple 
the spin temperature with the gas temperature. 
The differences among the models become remarkable at lower redshifts. 
The strength of Ly$\alpha$ coupling mainly determines the depth of the absorption because the ionized  fraction is still small in all the models. 
Since the LW background intensity is larger in Run-Fesc05 as shown in Fig.~\ref{fig:J21fesc}, less massive haloes cannot host stars so that the coupling is weaker than the other models. 
The differences among Run-Ms500, Run-Ms200, and Run-Ms80 are originated from the differences of $J_\mathrm{LW}$ as well: 
The absorption signal becomes deeper as the stellar mass decreases, because the escape fraction is low in the case of less massive Pop III star. 

The effect of UV heating is notable only if the ionization fraction is roughly more than $1\%$. 
In Run-Fesc05, the global signal with UV heating diverges from that without the heating around $z\sim22$, and the difference becomes larger as the ionization fraction increases. 
In the other runs of Run-Ms500, Run-Ms200, and Run-Ms80, however, UV heating does not have significant effect because the ionization fraction remains tiny. 

\begin{figure}
    \includegraphics[width=\columnwidth]{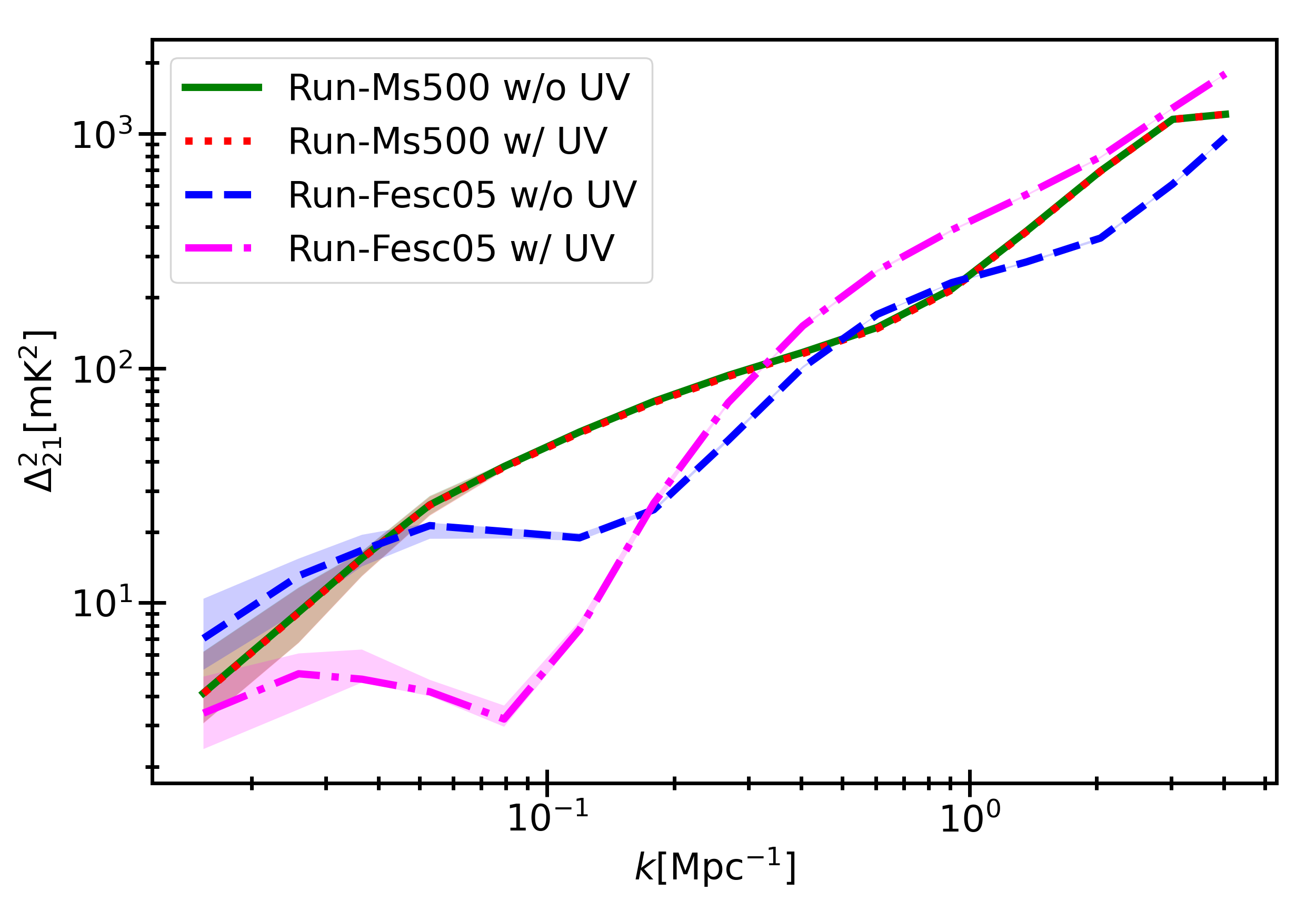}
    \caption{
    The power spectra of the 21-cm brightness temperature at redshift 20. 
    The green solid and red dotted curves respectively indicate the results of Run-Ms500 without and with UV heating. 
    The blue dashed and magenta dashed-dotted curves are thoese of Run-Fesc05.
    The shaded regions correspond to the 10 - 90 percentiles obtained from 10 realizations. }
    \label{fig:ps}
\end{figure}

Fig.~\ref{fig:ps} shows the power spectra of 21-cm brightness temperature,
$\Delta^2_{21}(k,z) \equiv \overline{\delta T}_\mathrm{b}^2 (z)\,\langle | \delta_{21}(\mathbf{k},z) |^2  \rangle k^3/(2\pi V)$,  at $z=20$, where $\delta_{21}(\mathbf{x},z) \equiv [ \delta T_\mathrm{b}(\mathbf{x},z) - \overline{\delta T}_\mathrm{b}(z)]/ \overline{\delta T}_\mathrm{b}(z)$. 
We first focus on the power spectrum obtained from Run-Ms500.  
In this case the power spectrum has relatively flat shape at the middle scale range ($k\sim 10^{-1} - 1$~[Mpc$^{-1}$]) and drops at larger scale ($k\lesssim 10^{-1}$~[Mpc$^{-1}$]). 
This trend mainly comes form the fluctuation of overdensity, and consistent with the results of \cite{Mesinger2011} when the Universe is neutral and Ly$\alpha$ coupling turns on.
Since the ionization fraction is small in Run-Ms500, UV heating does not affect the power spectrum as well as the 21-cm global signal.

The power spectrum in the Run-Fesc05 in which the ionization fraction is relatively large, on the other hand, slightly differs from that in Run-Ms500 due to the additional fluctuations of the neutral fraction and Ly$\alpha$ radiation. 
Moreover, UV heating  changes the shape of the power spectrum at large scales. From Fig.~\ref{fig:Tbmap}, we can actually see the 21-cm fluctuation is relatively small at large scales. 
This fact suggests that UV heating has non-negligible impact on the 21-cm power spectrum when the ionization fraction is $x_e \gtrsim 1\%$. 
Such relatively high ionization fraction can be achieved if the escape fraction is higher than our model. 
This point is discussed later in Section~\ref{sec:discussion}. 

\section{Discussion}
\label{sec:discussion}
In this work, we use the box-averaged LW intensity and single value of escape fraction at each redshift, however, the LW intensity can fluctuate largely \citep{Ahn2009},  which may have not a little impact on the escape fraction and the 21-cm signals. 
At high density regions, the LW intensity rises earlier than the average and the escape fraction falls toward zero, while, the low density region can remain low-intensity levels in which less massive haloes, whose escape fraction is large, can still form stars and possibly contribute to reionization. 
Thus, we incorporate the inhomogeneous LW feedback in the forthcoming paper.


The escape fraction used in our model is obtained from spherically symmetric 1D RHD simulations. 
It could be expected that multidimensional effects boost the escape fraction, because ionizing photons can escape more easily along directions with lower optical depth. 
If this is the case, UV heating likely has large impact on 21-cm signal even if the stellar mass and halo mass dependent escape fraction is considered. 
We here note that alternative escape fraction models can be easily incorporated by the method described in this paper.  
Thus, in our future works, we compute 21-signals with other escape fraction models. 

Since the main purpose of this paper is to develop the model of MHs hosting Pop III stars for semi-numerical simulations, we neglect X-ray heating. 
The Pop III stars would end up as high mass X-ray binaries (HMXBs) which are often presumed to be dominant X-ray source at high redshift \citep{Fragos2013}.
Such early X-ray photons can heat (and also ionize) the gas enough to largely affect the 21-cm signals \citep[e.g.][]{Fialkov2014}. 
Since the mean free path of X-ray is much longer than that of UV photons, the different fluctuation scale of gas heating would help us distinguish the two types of heating (see \citealt{Das2017} as example of different scale of 21-cm fluctuations originated from the mean free path difference). Thus, we are planing to investigate 21-cm power spectrum and its time evolution with both UV and X-ray heating in our future work.

\section{Conclusion}
\label{sec:conclusion}
We have developed a novel module of Pop III stars for semi-numerical simulations of 21-cm signal in the high-$z$ Universe. 
To model Pop III stars in detail, we first perform 1D RHD simulations.
By utilizing the RHD simulation results, we model the escape fractions of ionizing and LW photons considering the dependence on the halo mass and the stellar mass. 
We also take into account the photo-heating by Pop III stars, which has ever been neglected in the calculation of the cosmological 21-cm signal. 

We have then implemented the module into a public 21-cm semi-numerical simulation code  {\small 21CMFAST} and have computed the 21-cm signals. As a result, we have found that the contribution of Pop III stars to cosmic reionization strongly depends on the treatment of the escape fractions. 
The LW feedback significantly suppresses the Pop III star formation in less massive halos. 
Therefore with the halo-mass-dependent escape fraction, in which the escape fractions are higher for less massive halos, Pop III stars hardly contribute to reionization. 
On the other hand, the constant escape fraction which is conventionally assumed in simulations of 21-cm signals, enables Pop III stars to provide the IGM with enough amount of ionizing photons. 

We also found that the impact of the photo-heating by UV radiation on the 21-cm global signal appears when the ionization fraction is greater than $\sim 1\%$. 
If the Universe is hardly ionized, the absorption feature of the global signal is determined by the strength of the Ly$\alpha$ coupling (star formation rate density). 
The power spectrum in this case is almost controlled by the overdensity. 
On the other hand, in the case that the ionization fraction is as high as $\sim 1\%$, UV radiation mildly heats the IGM and the depth of the absorption signal becomes shallow. 
The heating also produces an additional feature on power spectra. 

Such difference is mainly caused by the model of the escape fraction. 
Although Pop III stars do not contribute to reionization with our escape fraction model, the model still could have uncertainties. 
If the escape fraction is relatively higher than our model, the efficiency of the heating depends on the stellar mass because of the stellar-mass-dependent escape fraction. 
In this case, we possibly extract the information of Pop III star mass from the 21-cm signal. In conclusion, detailed modelling of Pop III stars is essential to use the forthcoming 21-cm data beneficially.

\section*{Acknowledgements}
We are grateful to Tomoaki Ishiyama for kindly providing the N-body simulation data. The simulations in this work were performed with a cluster ‘galaxy’ installed in Nagoya University and Cray XC50 at Center for Computational Astrophysics, National Astronomical Observatory of Japan. 
This work was supported by JSPS KAKENHI Grant Number JP20J13423 (TT), JP18K03699 (KH).

\section*{DATA AVAILABILITY}
The data underlying this article will be shared on reasonable request to the corresponding author.




\bibliographystyle{mnras}
\bibliography{example} 

\begin{thebibliography}{}
\makeatletter
\relax
\def\mn@urlcharsother{\let\do\@makeother \do\$\do\&\do\#\do\^\do\_\do\%\do\~}
\def\mn@doi{\begingroup\mn@urlcharsother \@ifnextchar [ {\mn@doi@}
  {\mn@doi@[]}}
\def\mn@doi@[#1]#2{\def\@tempa{#1}\ifx\@tempa\@empty \href
  {http://dx.doi.org/#2} {doi:#2}\else \href {http://dx.doi.org/#2} {#1}\fi
  \endgroup}
\def\mn@eprint#1#2{\mn@eprint@#1:#2::\@nil}
\def\mn@eprint@arXiv#1{\href {http://arxiv.org/abs/#1} {{\tt arXiv:#1}}}
\def\mn@eprint@dblp#1{\href {http://dblp.uni-trier.de/rec/bibtex/#1.xml}
  {dblp:#1}}
\def\mn@eprint@#1:#2:#3:#4\@nil{\def\@tempa {#1}\def\@tempb {#2}\def\@tempc
  {#3}\ifx \@tempc \@empty \let \@tempc \@tempb \let \@tempb \@tempa \fi \ifx
  \@tempb \@empty \def\@tempb {arXiv}\fi \@ifundefined
  {mn@eprint@\@tempb}{\@tempb:\@tempc}{\expandafter \expandafter \csname
  mn@eprint@\@tempb\endcsname \expandafter{\@tempc}}}

\bibitem[\protect\citeauthoryear{{Abel}, {Bryan}  \& {Norman}}{{Abel}
  et~al.}{2002}]{Abel2002}
{Abel} T.,  {Bryan} G.~L.,   {Norman} M.~L.,  2002, \mn@doi [Science]
  {10.1126/science.295.5552.93}, \href
  {http://adsabs.harvard.edu/abs/2002Sci...295...93A} {295, 93}

\bibitem[\protect\citeauthoryear{{Ahn}, {Shapiro}, {Iliev}, {Mellema}  \&
  {Pen}}{{Ahn} et~al.}{2009}]{Ahn2009}
{Ahn} K.,  {Shapiro} P.~R.,  {Iliev} I.~T.,  {Mellema} G.,   {Pen} U.-L.,
  2009, \mn@doi [\apj] {10.1088/0004-637X/695/2/1430}, \href
  {https://ui.adsabs.harvard.edu/abs/2009ApJ...695.1430A} {695, 1430}

\bibitem[\protect\citeauthoryear{{Ahn}, {Iliev}, {Shapiro}, {Mellema}, {Koda}
  \& {Mao}}{{Ahn} et~al.}{2012}]{Ahn2012}
{Ahn} K.,  {Iliev} I.~T.,  {Shapiro} P.~R.,  {Mellema} G.,  {Koda} J.,   {Mao}
  Y.,  2012, \mn@doi [\apjl] {10.1088/2041-8205/756/1/L16}, \href
  {https://ui.adsabs.harvard.edu/abs/2012ApJ...756L..16A} {756, L16}

\bibitem[\protect\citeauthoryear{{Alvarez}, {Bromm}  \& {Shapiro}}{{Alvarez}
  et~al.}{2006}]{Alvarez2006}
{Alvarez} M.~A.,  {Bromm} V.,   {Shapiro} P.~R.,  2006, \mn@doi [\apj]
  {10.1086/499578}, \href {http://adsabs.harvard.edu/abs/2006ApJ...639..621A}
  {639, 621}

\bibitem[\protect\citeauthoryear{{Ba{\~n}ados} et~al.,}{{Ba{\~n}ados}
  et~al.}{2014}]{Banados2014}
{Ba{\~n}ados} E.,  et~al., 2014, \mn@doi [\aj] {10.1088/0004-6256/148/1/14},
  \href {http://adsabs.harvard.edu/abs/2014AJ....148...14B} {148, 14}

\bibitem[\protect\citeauthoryear{{Barkana} \& {Loeb}}{{Barkana} \&
  {Loeb}}{2005}]{Barkana2005}
{Barkana} R.,  {Loeb} A.,  2005, \mn@doi [\apj] {10.1086/429954}, \href
  {https://ui.adsabs.harvard.edu/abs/2005ApJ...626....1B} {626, 1}

\bibitem[\protect\citeauthoryear{{Barry} et~al.,}{{Barry}
  et~al.}{2019}]{Barry2019}
{Barry} N.,  et~al., 2019, \mn@doi [\apj] {10.3847/1538-4357/ab40a8}, \href
  {https://ui.adsabs.harvard.edu/abs/2019ApJ...884....1B} {884, 1}

\bibitem[\protect\citeauthoryear{{Beardsley} et~al.,}{{Beardsley}
  et~al.}{2016}]{Beardsley2016}
{Beardsley} A.~P.,  et~al., 2016, \mn@doi [\apj] {10.3847/1538-4357/833/1/102},
  \href {https://ui.adsabs.harvard.edu/abs/2016ApJ...833..102B} {833, 102}

\bibitem[\protect\citeauthoryear{{Bowman}, {Rogers}, {Monsalve}, {Mozdzen}  \&
  {Mahesh}}{{Bowman} et~al.}{2018}]{Bowman2018}
{Bowman} J.~D.,  {Rogers} A.~E.~E.,  {Monsalve} R.~A.,  {Mozdzen} T.~J.,
  {Mahesh} N.,  2018, \mn@doi [\nat] {10.1038/nature25792}, \href
  {http://adsabs.harvard.edu/abs/2018Natur.555...67B} {555, 67}

\bibitem[\protect\citeauthoryear{{Bromm}, {Coppi}  \& {Larson}}{{Bromm}
  et~al.}{2002}]{Bromm2002}
{Bromm} V.,  {Coppi} P.~S.,   {Larson} R.~B.,  2002, \mn@doi [\apj]
  {10.1086/323947}, \href {http://adsabs.harvard.edu/abs/2002ApJ...564...23B}
  {564, 23}

\bibitem[\protect\citeauthoryear{{Chen} \& {Miralda-Escud{\'e}}}{{Chen} \&
  {Miralda-Escud{\'e}}}{2008}]{Chen2008}
{Chen} X.,  {Miralda-Escud{\'e}} J.,  2008, \mn@doi [\apj] {10.1086/528941},
  \href {http://adsabs.harvard.edu/abs/2008ApJ...684...18C} {684, 18}

\bibitem[\protect\citeauthoryear{{Das}, {Mesinger}, {Pallottini}, {Ferrara}  \&
  {Wise}}{{Das} et~al.}{2017}]{Das2017}
{Das} A.,  {Mesinger} A.,  {Pallottini} A.,  {Ferrara} A.,   {Wise} J.~H.,
  2017, \mn@doi [\mnras] {10.1093/mnras/stx943}, \href
  {https://ui.adsabs.harvard.edu/abs/2017MNRAS.469.1166D} {469, 1166}

\bibitem[\protect\citeauthoryear{{Dillon} et~al.,}{{Dillon}
  et~al.}{2014}]{Dillon2014}
{Dillon} J.~S.,  et~al., 2014, \mn@doi [\prd] {10.1103/PhysRevD.89.023002},
  \href {https://ui.adsabs.harvard.edu/abs/2014PhRvD..89b3002D} {89, 023002}

\bibitem[\protect\citeauthoryear{{Dillon} et~al.,}{{Dillon}
  et~al.}{2015}]{Dillon2015}
{Dillon} J.~S.,  et~al., 2015, \mn@doi [\prd] {10.1103/PhysRevD.91.123011},
  \href {https://ui.adsabs.harvard.edu/abs/2015PhRvD..91l3011D} {91, 123011}

\bibitem[\protect\citeauthoryear{{Eastwood} et~al.,}{{Eastwood}
  et~al.}{2019}]{Eastwood2019}
{Eastwood} M.~W.,  et~al., 2019, \mn@doi [\aj] {10.3847/1538-3881/ab2629},
  \href {https://ui.adsabs.harvard.edu/abs/2019AJ....158...84E} {158, 84}

\bibitem[\protect\citeauthoryear{{Ewall-Wice} et~al.,}{{Ewall-Wice}
  et~al.}{2016}]{Ewall-Wice2016}
{Ewall-Wice} A.,  et~al., 2016, \mn@doi [\mnras] {10.1093/mnras/stw1022}, \href
  {https://ui.adsabs.harvard.edu/abs/2016MNRAS.460.4320E} {460, 4320}

\bibitem[\protect\citeauthoryear{{Fialkov}, {Barkana}, {Visbal},
  {Tseliakhovich}  \& {Hirata}}{{Fialkov} et~al.}{2013}]{Fialkov2013}
{Fialkov} A.,  {Barkana} R.,  {Visbal} E.,  {Tseliakhovich} D.,   {Hirata}
  C.~M.,  2013, \mn@doi [\mnras] {10.1093/mnras/stt650}, \href
  {https://ui.adsabs.harvard.edu/abs/2013MNRAS.432.2909F} {432, 2909}

\bibitem[\protect\citeauthoryear{{Fialkov}, {Barkana}  \& {Visbal}}{{Fialkov}
  et~al.}{2014}]{Fialkov2014}
{Fialkov} A.,  {Barkana} R.,   {Visbal} E.,  2014, \mn@doi [\nat]
  {10.1038/nature12999}, \href
  {https://ui.adsabs.harvard.edu/abs/2014Natur.506..197F} {506, 197}

\bibitem[\protect\citeauthoryear{{Fragos} et~al.,}{{Fragos}
  et~al.}{2013}]{Fragos2013}
{Fragos} T.,  et~al., 2013, \mn@doi [\apj] {10.1088/0004-637X/764/1/41}, \href
  {https://ui.adsabs.harvard.edu/abs/2013ApJ...764...41F} {764, 41}

\bibitem[\protect\citeauthoryear{{Furlanetto} \& {Oh}}{{Furlanetto} \&
  {Oh}}{2006}]{Furlanetto2006Oh}
{Furlanetto} S.~R.,  {Oh} S.~P.,  2006, \mn@doi [\apj] {10.1086/508448}, \href
  {http://adsabs.harvard.edu/abs/2006ApJ...652..849F} {652, 849}

\bibitem[\protect\citeauthoryear{{Gao}, {Yoshida}, {Abel}, {Frenk}, {Jenkins}
  \& {Springel}}{{Gao} et~al.}{2007}]{Gao2007}
{Gao} L.,  {Yoshida} N.,  {Abel} T.,  {Frenk} C.~S.,  {Jenkins} A.,
  {Springel} V.,  2007, \mn@doi [\mnras] {10.1111/j.1365-2966.2007.11814.x},
  \href {http://adsabs.harvard.edu/abs/2007MNRAS.378..449G} {378, 449}

\bibitem[\protect\citeauthoryear{{Gehlot} et~al.,}{{Gehlot}
  et~al.}{2019}]{Gehlot2019}
{Gehlot} B.~K.,  et~al., 2019, \mn@doi [\mnras] {10.1093/mnras/stz1937}, \href
  {https://ui.adsabs.harvard.edu/abs/2019MNRAS.488.4271G} {488, 4271}

\bibitem[\protect\citeauthoryear{{Haiman}, {Rees}  \& {Loeb}}{{Haiman}
  et~al.}{1997}]{Haiman1997}
{Haiman} Z.,  {Rees} M.~J.,   {Loeb} A.,  1997, \mn@doi [\apj]
  {10.1086/304386}, \href
  {https://ui.adsabs.harvard.edu/abs/1997ApJ...484..985H} {484, 985}

\bibitem[\protect\citeauthoryear{{Hasegawa}, {Umemura}  \& {Susa}}{{Hasegawa}
  et~al.}{2009}]{Hasegawa2009}
{Hasegawa} K.,  {Umemura} M.,   {Susa} H.,  2009, \mn@doi [\mnras]
  {10.1111/j.1365-2966.2009.14639.x}, \href
  {https://ui.adsabs.harvard.edu/abs/2009MNRAS.395.1280H} {395, 1280}

\bibitem[\protect\citeauthoryear{{Hirano}, {Hosokawa}, {Yoshida}, {Omukai}  \&
  {Yorke}}{{Hirano} et~al.}{2015}]{Hirano2015}
{Hirano} S.,  {Hosokawa} T.,  {Yoshida} N.,  {Omukai} K.,   {Yorke} H.~W.,
  2015, \mn@doi [\mnras] {10.1093/mnras/stv044}, \href
  {http://adsabs.harvard.edu/abs/2015MNRAS.448..568H} {448, 568}

\bibitem[\protect\citeauthoryear{{Ishiyama}, {Sudo}, {Yokoi}, {Hasegawa},
  {Tominaga}  \& {Susa}}{{Ishiyama} et~al.}{2016}]{Ishiyama2016}
{Ishiyama} T.,  {Sudo} K.,  {Yokoi} S.,  {Hasegawa} K.,  {Tominaga} N.,
  {Susa} H.,  2016, \mn@doi [\apj] {10.3847/0004-637X/826/1/9}, \href
  {https://ui.adsabs.harvard.edu/abs/2016ApJ...826....9I} {826, 9}

\bibitem[\protect\citeauthoryear{{Johnson}, {Greif}  \& {Bromm}}{{Johnson}
  et~al.}{2007}]{Johnson2007}
{Johnson} J.~L.,  {Greif} T.~H.,   {Bromm} V.,  2007, \mn@doi [\apj]
  {10.1086/519212}, \href {http://adsabs.harvard.edu/abs/2007ApJ...665...85J}
  {665, 85}

\bibitem[\protect\citeauthoryear{{Karlsson}, {Bromm}  \&
  {Bland-Hawthorn}}{{Karlsson} et~al.}{2013}]{Karlsson2013}
{Karlsson} T.,  {Bromm} V.,   {Bland-Hawthorn} J.,  2013, \mn@doi [Reviews of
  Modern Physics] {10.1103/RevModPhys.85.809}, \href
  {http://adsabs.harvard.edu/abs/2013RvMP...85..809K} {85, 809}

\bibitem[\protect\citeauthoryear{{Kitayama}, {Yoshida}, {Susa}  \&
  {Umemura}}{{Kitayama} et~al.}{2004}]{Kitayama2004}
{Kitayama} T.,  {Yoshida} N.,  {Susa} H.,   {Umemura} M.,  2004, \mn@doi [\apj]
  {10.1086/423313}, \href {http://adsabs.harvard.edu/abs/2004ApJ...613..631K}
  {613, 631}

\bibitem[\protect\citeauthoryear{{Kolopanis} et~al.,}{{Kolopanis}
  et~al.}{2019}]{Kolopanis2019}
{Kolopanis} M.,  et~al., 2019, \mn@doi [\apj] {10.3847/1538-4357/ab3e3a}, \href
  {https://ui.adsabs.harvard.edu/abs/2019ApJ...883..133K} {883, 133}

\bibitem[\protect\citeauthoryear{{Machacek}, {Bryan}  \& {Abel}}{{Machacek}
  et~al.}{2001}]{Machacek2001}
{Machacek} M.~E.,  {Bryan} G.~L.,   {Abel} T.,  2001, \mn@doi [\apj]
  {10.1086/319014}, \href
  {https://ui.adsabs.harvard.edu/abs/2001ApJ...548..509M} {548, 509}

\bibitem[\protect\citeauthoryear{{Mebane}, {Mirocha}  \& {Furlanetto}}{{Mebane}
  et~al.}{2018}]{Mebane2018}
{Mebane} R.~H.,  {Mirocha} J.,   {Furlanetto} S.~R.,  2018, \mn@doi [\mnras]
  {10.1093/mnras/sty1833}, \href
  {https://ui.adsabs.harvard.edu/abs/2018MNRAS.479.4544M} {479, 4544}

\bibitem[\protect\citeauthoryear{{Mesinger}, {Furlanetto}  \& {Cen}}{{Mesinger}
  et~al.}{2011}]{Mesinger2011}
{Mesinger} A.,  {Furlanetto} S.,   {Cen} R.,  2011, \mn@doi [\mnras]
  {10.1111/j.1365-2966.2010.17731.x}, \href
  {https://ui.adsabs.harvard.edu/abs/2011MNRAS.411..955M} {411, 955}

\bibitem[\protect\citeauthoryear{{Nishi} \& {Susa}}{{Nishi} \&
  {Susa}}{1999}]{Nishi1999}
{Nishi} R.,  {Susa} H.,  1999, \mn@doi [\apjl] {10.1086/312277}, \href
  {http://adsabs.harvard.edu/abs/1999ApJ...523L.103N} {523, L103}

\bibitem[\protect\citeauthoryear{{O'Shea} \& {Norman}}{{O'Shea} \&
  {Norman}}{2007}]{O'Shea2007}
{O'Shea} B.~W.,  {Norman} M.~L.,  2007, \mn@doi [\apj] {10.1086/509250}, \href
  {http://adsabs.harvard.edu/abs/2007ApJ...654...66O} {654, 66}

\bibitem[\protect\citeauthoryear{{O'Shea} \& {Norman}}{{O'Shea} \&
  {Norman}}{2008}]{OShea2008}
{O'Shea} B.~W.,  {Norman} M.~L.,  2008, \mn@doi [\apj] {10.1086/524006}, \href
  {https://ui.adsabs.harvard.edu/abs/2008ApJ...673...14O} {673, 14}

\bibitem[\protect\citeauthoryear{{Paciga} et~al.,}{{Paciga}
  et~al.}{2013}]{Paciga2013}
{Paciga} G.,  et~al., 2013, \mn@doi [\mnras] {10.1093/mnras/stt753}, \href
  {https://ui.adsabs.harvard.edu/abs/2013MNRAS.433..639P} {433, 639}

\bibitem[\protect\citeauthoryear{{Patil} et~al.,}{{Patil}
  et~al.}{2017}]{Patil2017}
{Patil} A.~H.,  et~al., 2017, \mn@doi [\apj] {10.3847/1538-4357/aa63e7}, \href
  {https://ui.adsabs.harvard.edu/abs/2017ApJ...838...65P} {838, 65}

\bibitem[\protect\citeauthoryear{{Planck Collaboration} et~al.,}{{Planck
  Collaboration} et~al.}{2016}]{PlanckXIII}
{Planck Collaboration} et~al., 2016, \mn@doi [\aap]
  {10.1051/0004-6361/201525830}, \href
  {http://adsabs.harvard.edu/abs/2016A%26A...594A..13P} {594, A13}

\bibitem[\protect\citeauthoryear{{Schaerer}}{{Schaerer}}{2002}]{Schaerer2002}
{Schaerer} D.,  2002, \mn@doi [\aap] {10.1051/0004-6361:20011619}, \href
  {http://adsabs.harvard.edu/abs/2002A\%26A...382...28S} {382, 28}

\bibitem[\protect\citeauthoryear{{Shapiro} et~al.,}{{Shapiro}
  et~al.}{2012}]{Shapiro2012}
{Shapiro} P.~R.,  et~al., 2012, in {Umemura} M.,  {Omukai} K.,  eds,  American
  Institute of Physics Conference Series Vol. 1480, American Institute of
  Physics Conference Series. pp 248--260 (\mn@eprint {arXiv} {1211.0583}),
  \mn@doi{10.1063/1.4754363}

\bibitem[\protect\citeauthoryear{{Sobacchi} \& {Mesinger}}{{Sobacchi} \&
  {Mesinger}}{2014}]{Sobacchi2014}
{Sobacchi} E.,  {Mesinger} A.,  2014, \mn@doi [\mnras] {10.1093/mnras/stu377},
  \href {https://ui.adsabs.harvard.edu/abs/2014MNRAS.440.1662S} {440, 1662}

\bibitem[\protect\citeauthoryear{{Susa} \& {Umemura}}{{Susa} \&
  {Umemura}}{2006}]{Susa2006}
{Susa} H.,  {Umemura} M.,  2006, \mn@doi [\apjl] {10.1086/506275}, \href
  {https://ui.adsabs.harvard.edu/abs/2006ApJ...645L..93S} {645, L93}

\bibitem[\protect\citeauthoryear{{Susa}, {Hasegawa}  \& {Tominaga}}{{Susa}
  et~al.}{2014}]{Susa2014}
{Susa} H.,  {Hasegawa} K.,   {Tominaga} N.,  2014, \mn@doi [\apj]
  {10.1088/0004-637X/792/1/32}, \href
  {http://adsabs.harvard.edu/abs/2014ApJ...792...32S} {792, 32}

\bibitem[\protect\citeauthoryear{{Tanaka}, {Hasegawa}, {Yajima}, {Kobayashi}
  \& {Sugiyama}}{{Tanaka} et~al.}{2018}]{TT2018}
{Tanaka} T.,  {Hasegawa} K.,  {Yajima} H.,  {Kobayashi} M. I.~N.,   {Sugiyama}
  N.,  2018, \mn@doi [\mnras] {10.1093/mnras/sty1967}, \href
  {https://ui.adsabs.harvard.edu/abs/2018MNRAS.480.1925T} {480, 1925}

\bibitem[\protect\citeauthoryear{{Tegmark}, {Silk}, {Rees}, {Blanchard}, {Abel}
   \& {Palla}}{{Tegmark} et~al.}{1997}]{Tegmark1997}
{Tegmark} M.,  {Silk} J.,  {Rees} M.~J.,  {Blanchard} A.,  {Abel} T.,   {Palla}
  F.,  1997, \mn@doi [\apj] {10.1086/303434}, \href
  {http://adsabs.harvard.edu/abs/1997ApJ...474....1T} {474, 1}

\bibitem[\protect\citeauthoryear{{Venemans} et~al.,}{{Venemans}
  et~al.}{2013}]{Venemans2013}
{Venemans} B.~P.,  et~al., 2013, \mn@doi [\apj] {10.1088/0004-637X/779/1/24},
  \href {http://adsabs.harvard.edu/abs/2013ApJ...779...24V} {779, 24}

\bibitem[\protect\citeauthoryear{{Visbal}, {Haiman}, {Terrazas}, {Bryan}  \&
  {Barkana}}{{Visbal} et~al.}{2014}]{Visbal2014}
{Visbal} E.,  {Haiman} Z.,  {Terrazas} B.,  {Bryan} G.~L.,   {Barkana} R.,
  2014, \mn@doi [\mnras] {10.1093/mnras/stu1710}, \href
  {https://ui.adsabs.harvard.edu/abs/2014MNRAS.445..107V} {445, 107}

\bibitem[\protect\citeauthoryear{{Visbal}, {Haiman}  \& {Bryan}}{{Visbal}
  et~al.}{2018}]{Visbal2018}
{Visbal} E.,  {Haiman} Z.,   {Bryan} G.~L.,  2018, \mn@doi [\mnras]
  {10.1093/mnras/sty142}, \href
  {https://ui.adsabs.harvard.edu/abs/2018MNRAS.475.5246V} {475, 5246}

\bibitem[\protect\citeauthoryear{{Visbal}, {Bryan}  \& {Haiman}}{{Visbal}
  et~al.}{2020}]{Visbal2020}
{Visbal} E.,  {Bryan} G.~L.,   {Haiman} Z.,  2020, \mn@doi [\apj]
  {10.3847/1538-4357/ab994e}, \href
  {https://ui.adsabs.harvard.edu/abs/2020ApJ...897...95V} {897, 95}

\bibitem[\protect\citeauthoryear{{Wise} \& {Abel}}{{Wise} \&
  {Abel}}{2007}]{Wise2007}
{Wise} J.~H.,  {Abel} T.,  2007, \mn@doi [\apj] {10.1086/522876}, \href
  {https://ui.adsabs.harvard.edu/abs/2007ApJ...671.1559W} {671, 1559}

\bibitem[\protect\citeauthoryear{{Wise}, {Turk}, {Norman}  \& {Abel}}{{Wise}
  et~al.}{2012}]{Wise2012}
{Wise} J.~H.,  {Turk} M.~J.,  {Norman} M.~L.,   {Abel} T.,  2012, \mn@doi
  [\apj] {10.1088/0004-637X/745/1/50}, \href
  {http://adsabs.harvard.edu/abs/2012ApJ...745...50W} {745, 50}

\bibitem[\protect\citeauthoryear{{Wu} et~al.,}{{Wu} et~al.}{2015}]{Wu2015}
{Wu} X.-B.,  et~al., 2015, \mn@doi [\nat] {10.1038/nature14241}, \href
  {http://adsabs.harvard.edu/abs/2015Natur.518..512W} {518, 512}

\bibitem[\protect\citeauthoryear{{Yajima} \& {Li}}{{Yajima} \&
  {Li}}{2014}]{Yajima2014}
{Yajima} H.,  {Li} Y.,  2014, \mn@doi [\mnras] {10.1093/mnras/stu1982}, \href
  {http://adsabs.harvard.edu/abs/2014MNRAS.445.3674Y} {445, 3674}

\bibitem[\protect\citeauthoryear{{Yoshida}, {Omukai}  \& {Hernquist}}{{Yoshida}
  et~al.}{2008}]{Yoshida2008}
{Yoshida} N.,  {Omukai} K.,   {Hernquist} L.,  2008, \mn@doi [Science]
  {10.1126/science.1160259}, \href
  {http://adsabs.harvard.edu/abs/2008Sci...321..669Y} {321, 669}

\makeatother
\end{thebibliography}






\bsp	
\label{lastpage}
\end{document}